\documentclass[10pt]{article}
\usepackage{graphicx} % Required for inserting images
\usepackage{amsmath}
\usepackage{lipsum}
\usepackage{pdfcomment}

\usepackage[table,xcdraw]{xcolor} % Enable table colors and HTML color codes
\usepackage{subcaption}
\usepackage{hyperref} % Include the hyperref package
\usepackage[english]{babel}
\usepackage[utf8]{inputenc}
\usepackage{amsfonts}
\usepackage{longtable}
\usepackage{booktabs}
\usepackage{float} 
\usepackage{geometry}
\usepackage{lineno}
\usepackage{authblk}
\usepackage{tcolorbox}
\usepackage{rotating}
\usepackage{booktabs}
\usepackage{threeparttable}
\usepackage{geometry} % Optional: for full page layout control
\usepackage{lscape}
\usepackage{siunitx}  % For aligning numbers on the decimal point
\usepackage{adjustbox} % To adjust table width if needed

\usepackage{cite}

\usepackage{mathpazo}

\usepackage{array}
\usepackage{multirow}

\usepackage{amssymb}
\usepackage{pifont}

\usepackage[colorinlistoftodos]{todonotes}

\usepackage{wrapfig}

\usepackage[numbers,sort]{natbib}
\usepackage{cite}

\hypersetup{
    colorlinks=true, % Enable colored links
    linkcolor=black,  % Set the color of internal links (e.g., references)
    citecolor=blue, % Change this color as desired
    urlcolor=blue    % Set the color of external links (e.g., URLs)
}

\usepackage{amsthm}
\usepackage{titlesec}
\usepackage{titletoc}
\titlecontents{subsubsubsection}
[8em] % adjust this value to fit your needs
{\small}
{\hyperlink{toc.\thecontentslabel}{\thecontentslabel.} }
{}
{\ \titlerule*[.5pc]{.}\contentspage}

\titlespacing*{\subsubsubsection}{0pt}{3.25ex plus 1ex minus .2ex}{1.5ex plus .2ex}

\usepackage{caption}
\usepackage{microtype}

\usepackage{enumitem}
\setlist{itemsep=0em}

\usepackage{fancyhdr} % For custom headers/footers
\usepackage{lastpage} % For total number of pages

\usepackage{listings}
\usepackage{xcolor}

\definecolor{codegreen}{rgb}{0,0.6,0}
\definecolor{codegray}{rgb}{0.5,0.5,0.5}
\definecolor{codepurple}{rgb}{0.58,0,0.82}
\definecolor{backcolour}{rgb}{0.95,0.95,0.92}

\lstdefinestyle{mystyle}{
    backgroundcolor=\color{backcolour},   
    commentstyle=\color{codegreen},
    keywordstyle=\color{magenta},
    numberstyle=\tiny\color{codegray},
    stringstyle=\color{codepurple},
    basicstyle=\ttfamily\footnotesize,
    breakatwhitespace=false,         
    breaklines=true,                 
    captionpos=b,                    
    keepspaces=true,                 
    numbers=left,                    
    numbersep=5pt,                  
    showspaces=false,                
    showstringspaces=false,
    showtabs=false,                  
    tabsize=2
}

\begin{document}

%=================================================================
% Full title of the paper (Capitalized)

\begin{center}
{\LARGE\textbf{Large Language Model-Driven Code Compliance Checking in Building Information Modeling}}
\end{center}

\vspace{0.5cm}      % Adds 1 centimeter of vertical space

% \date{}
% \author{}

% \maketitle    
{\small
\begin{center}
\begin{minipage}{0.32\linewidth}
\centering
\textbf{Soumya Madireddy} \\
\textit{Department of Civil and Environmental Engineering} \\
\textit{University of Houston} \\
Houston, USA \\
\texttt{smadire3@cougarnet.uh.edu}
\end{minipage}
\hfill
\begin{minipage}{0.32\linewidth}
\centering
\textbf{Lu Gao, Ph.D.} \\
\textit{Department of Civil and Environmental Engineering} \\
\textit{University of Houston} \\
Houston, USA \\
\texttt{lgao5@central.uh.edu}
\end{minipage}
\hfill
\begin{minipage}{0.32\linewidth}
\centering
\textbf{Zia Din, Ph.D.} \\
\textit{Department of Civil and Environmental Engineering} \\
\textit{University of Houston} \\
Houston, USA \\
\texttt{uziauddi@central.uh.edu}
\end{minipage}

\vspace{1.5em}

\begin{minipage}{0.32\linewidth}
\centering
\textbf{Kinam Kim, Ph.D.} \\
\textit{Department of Civil and Environmental Engineering} \\
\textit{University of Houston} \\
Houston, USA \\
\texttt{kkim48@central.uh.edu}
\end{minipage}
\hfill
\begin{minipage}{0.32\linewidth}
\centering
\textbf{Ahmed Senouci, Ph.D.} \\
\textit{Department of Civil and Environmental Engineering} \\
\textit{University of Houston} \\
Houston, USA \\
\texttt{asenouci@central.uh.edu}
\end{minipage}
\hfill
\begin{minipage}{0.32\linewidth}
\centering
\textbf{Zhe Han, Ph.D.} \\
\textit{Center for Transportation Research} \\
\textit{The University of Texas at Austin} \\
Austin, USA \\
\texttt{hanzhe@utexas.edu}
\end{minipage}

\vspace{1.5em}

\begin{minipage}{0.32\linewidth}
\centering
\textbf{Yunpeng Zhang, Ph.D.} \\
\textit{Department of Information Science Technology} \\
\textit{University of Houston} \\
Houston, USA \\
\texttt{yzhan226@central.uh.edu}
\end{minipage}
\end{center}
}

% Keywords

% Abstract (Do not insert blank lines, i.e. \\) 
\section*{Abstract}
This research addresses the time-consuming and error-prone nature of manual code compliance checking in Building Information Modeling (BIM) by introducing a Large Language Model (LLM)-driven approach to semi-automate this critical process. The developed system integrates LLMs such as GPT, Claude, Gemini, and Llama, with Revit software to interpret building codes, generate Python scripts, and perform semi-automated compliance checks within the BIM environment. Case studies on a single-family residential project and an office building project demonstrated the system's ability to reduce the time and effort required for compliance checks while improving accuracy. It streamlined the identification of violations, such as non-compliant room dimensions, material usage, and object placements, by automatically assessing relationships and generating actionable reports. Compared to manual methods, the system eliminated repetitive tasks, simplified complex regulations, and ensured reliable adherence to standards. By offering a comprehensive, adaptable, and cost-effective solution, this proposed approach offers a promising advancement in BIM-based compliance checking, with potential applications across diverse regulatory documents in construction projects.\\

% Keywords
\noindent \textbf{Keywords}:  BIM, Semi-Automated Compliance Checking, Code Compliance Checking, AI, LLM

\section{Introduction}

\subsection{Code Compliance}
Ensuring that construction projects adhere to building codes and regulations is a critical aspect of the architecture, engineering, and construction (AEC) industry \citep{zhang2023factors}. Traditionally, this process has been manual, relying on senior professionals to review designs using CAD drawings and specifications, which is time-consuming, inefficient, and prone to errors \citep{aydin2021building}. Automated Compliance Checking (ACC) has emerged as a solution to improve accuracy and efficiency by using computer programs to check building projects against codes. The use of Building Information Modeling (BIM) has enhanced ACC by providing a digital representation of a building, enabling better collaboration and information exchange \citep{ismail2017review}.

The core of ACC involves translating building codes and regulations into a machine-interpretable format \citep{preidel2015automated}. This process includes several key steps: rule interpretation, building model preparation, rule execution, and results reporting \citep{chen2024automated}. Rule interpretation is considered the most vital and complex stage, often involving techniques like using existing software, creating plug-in applications, or adopting object-based, logical, or ontological approaches \citep{ismail2017review}. The use of natural language processing (NLP) is also important for interpreting regulatory texts and converting them into computable rules \citep{wang2023automated}. The building model is prepared by extracting the necessary information from BIM models, including geometric and property data. During rule execution, the interpreted rules are applied to the prepared model, and any violations are recorded \citep{yogana2021development}.

These steps are increasingly supported by advanced digital technologies, such as BIM, NLP, and machine learning, which significantly improve the accuracy and speed of compliance checks \citep{bus2019towards}. For example, \citet{li2024automated} proposed an integrative framework for automated compliance checking in BIM models, utilizing knowledge graphs and NLP to identify errors based on building standards, specifically in architectural and fire safety contexts. \citet{lange2021strategically} developed a machine learning system to automate accessibility compliance checking in BIM designs; the system utilized a Convolutional Neural Network (CNN) to analyze BIM models and identify accessibility issues, such as urban surfaces or excessive ramp slopes with an accuracy of 95\%.

\subsection{Challenges in current ACC systems}

The current ACC systems face several persistent challenges that hinder their widespread implementation and effectiveness. A significant issue lies in translating the complex and often ambiguous language of building codes into a machine-interpretable format \citep{nawari2019blockchain}. Building codes frequently change, with new requirements added regularly, making it difficult to maintain and update these systems, especially when they rely on hard-coded rules \citep{preidel2017refinement}. Also, the interpretation of regulatory texts using NLP tools is complicated by the legal and technical nuances of the text \citep{zhao2023compliance}.

Another critical challenge is the reliance on detailed, accurate, and complete building information, as deficiencies in BIM models, such as missing or incorrect data, can significantly hinder the automated checking process \citep{bloch2020clustering}. The lack of standardization in BIM data further exacerbates these problems, often necessitating extensive manual preprocessing to correct inconsistencies and omissions \citep{de2024enriching}. Furthermore, existing ACC systems often operate as "black boxes," lacking transparency and flexibility, which makes them difficult to understand or modify for specific user needs \citep{preidel2015automated}. These systems also struggle with addressing qualitative aspects of building codes, such as aesthetics and spatial functionality, and face scalability issues as building designs become increasingly complex \citep{lee2023high}. 

Despite these challenges, ACC systems have demonstrated significant potential to improve the speed, accuracy, and consistency of code compliance checks. By leveraging advanced digital technologies such as BIM, NLP, and machine learning, these systems can save time and resources while reducing the likelihood of human error \citep{bus2019towards}. 

\subsection{Emergence of LLMs and its transformative potential}
The emergence of LLMs has introduced a transformative potential for ACC by offering advanced capabilities in natural language processing \citep{chen2024automated}. LLMs, pre-trained on vast amounts of data, demonstrate a strong ability to understand and generate human language with minimal task-specific training. This enables them to interpret complex regulatory texts and potentially convert them into computable rules more efficiently than previous methods \citep{zhang2023can}. LLMs can also adapt to new regulations and extract structured information from regulatory texts. LLMs also show promise in generating formal representations of regulations, potentially replacing the need for manual rule creation and improving the overall efficiency and effectiveness of checking processes \citep{fuchs2024using}. However, despite their great potential, there are still challenges to be addressed, such as ensuring the accuracy of the generated outputs and the dependence on prompt engineering to guide their responses effectively. These challenges show that more research is needed to make full use of LLMs for automated compliance checking.

\subsection{Objectives of this research}

While previous research has explored the application of LLMs to ACC, there are still challenges remain unresolved. Existing compliance-checking models lack integration with BIM environments. Many platforms are restricted to specific regulations and difficult to adapt to changes. To address this, we propose an LLM-based approach that converts regulations into executable Python scripts for real-time compliance checking in Revit.

\section{Literature Review}

\subsection{Importance of code compliance in construction}

Code compliance in construction is crucial for building safety, legal adherence, and efficiency through automation \citep{borrmann2018building}. Building codes are legal documents designed to protect public safety by establishing minimum standards for construction \citep{nguyen2011building}. They specify requirements for various aspects of building design and construction, such as fire safety, accessibility, and structural integrity \citep{lee2015automated}. By adhering to these codes, construction projects can avoid safety hazards that may lead to accidents, injuries, or even fatalities \citep{anderson2020using}. ACC systems play a crucial role in enhancing safety by reducing human error during the design, review, and construction planning phases \citep{zhang2023factors}. These systems can identify potential safety issues early in the design process, allowing for timely modifications and preventing costly and potentially dangerous rework. Moreover, automated systems can check for specific safety concerns like fall hazards and spatial relationships, ensuring a safer environment for both construction workers and future building occupants  \citep{anderson2020using}.

Code compliance also significantly impacts the efficiency of construction projects \citep{preidel2015automated}. Manual code checking is a time-consuming and error-prone process, often requiring extensive resources and leading to project delays and increased costs \citep{ismail2017review}. However, ACC systems offer a solution by streamlining the process, saving time, money, and labor \citep{anderson2020using}. BIM technology facilitates automated checking by providing a digital representation of building designs and related data, making it possible to perform checks more quickly and accurately \citep{zhao2023compliance}. By detecting and addressing potential issues during the design phase, ACC helps to prevent costly rework \citep{nguyen2011building}. Furthermore, ACC can streamline the building permit process, promote better collaboration among project participants, and produce inspection reports quickly \citep{lee2015automated}. 

Furthermore, code compliance is critical for legal adherence \citep{aydin2021building}. Building codes are legally binding and failure to comply can result in significant legal disputes, project delays, and financial penalties. Standardized codes and automated systems also promote greater consistency across jurisdictions, reducing confusion for designers and builders \citep{anderson2020using}.  A system that can perform checks with integrity and credibility is essential for maintaining trust and preventing legal problems in construction projects \citep{lee2023high}.

\subsection{Methods for ensuring compliance}
The development of methods for ensuring compliance in the AEC industry has progressed from manual, error-prone processes to sophisticated automated systems, driven by technological advancements, the increasing complexity of building codes, and the need for efficiency and accuracy \citep{anderson2020using}. Initially, compliance relied on manual interpretation and review of design drawings and specifications by experienced professionals \citep{yogana2021development}. This involved a time-consuming and costly process where senior personnel examined drawings, often repeating similar checks, and was highly susceptible to errors \citep{zhao2023compliance}. These manual checks were limited by the fact that professionals could not memorize all the codes, which were often contradictory \citep{zhang2023factors}. 

The introduction of computer-aided design (CAD) in the 1980s brought a transition to digital methods, but early CAD systems still relied on manual checking of drawings and textual descriptions, simply digitizing existing workflows. These systems lacked the ability to perform intelligent rule-based checks, and were essentially digital versions of the manual process  \citep{zhang2023factors}. BIM marked a significant shift in compliance checking by integrating building information with rule-based checking \citep{chen2024automated}. BIM is a digital model that captures the physical and functional characteristics of a facility. It acts as a shared knowledge resource to aid decision-making throughout a facility’s life \citep{pezeshki2018applications}. BIM technology has revolutionized the construction industry by improving collaboration, accuracy, and project management \citep{azhar2011building}. Initially focused on 3D modeling, BIM has expanded to incorporate 4D (time), 5D (cost), and 6D (sustainability) \citep{succar2010building}. BIM allowed for the creation of digital models containing not only the geometry of a building but also extensive data about its components and systems. Early BIM-based approaches used hard-coded rules directly into software, where compliance rules were embedded within the software's code \citep{zhu2025research}. These early systems were inflexible, difficult to modify, and often lacked transparency, acting as "black boxes" \citep{borrmann2018building}. 

To address the limitations of inflexibility and lack of transparency, semi-automated methods were introduced \citep{chen2024automated}. These methods aimed to translate regulatory text into machine-processable formats using logical operators, while still requiring some manual effort . 
The use of predicate logic allowed for the validation of checking methods, calculations, and conditions in the rules, expressing rules as logical conditions that could be evaluated. The deontological approach, using deontic logic, was introduced for more complex knowledge representation and reasoning \citep{salama2011semantic}.

Further advancements included the use of NLP techniques to automate the extraction of information from regulatory texts \citep{chen2024automated}. These techniques converted unstructured text into structured information for automated reasoning. Early NLP approaches included both rule-based and statistical methods, with rule-based methods generally offering better accuracy but requiring more human labor. Researchers explored methods to extract semantic and syntactic information and to categorize text in regulatory documents to improve efficiency \citep{zhang2015automated}. The development of ontologies also played a key role, enabling the representation of domain knowledge and supporting semantic reasoning \citep{zhong2018ontology}.

More recently, there has been a focus on utilizing visual programming languages to make rule-making and compliance checking more accessible to non-programmers. The Visual Code Checking Language (VCCL) allows users to visually translate codes and formalize them \citep{borrmann2018building}. Domain-specific languages, such as the Building Environment and Analysis Language (BERA), provide a mean to encode complex rules regarding spatial and circulation requirements \citep{preidel2015automated}.

The emergence of LLMs like ChatGPT has demonstrated the potential to automate the translation of natural language requirements into computable representations \citep{zhang2023can}. LLMs can address the limitations of deep learning by providing robust language understanding with minimal labeled data, adapting to evolving regulations, and accurately extracting structured information from regulatory texts  \citep{chen2024automated}.

Knowledge graphs are also being explored to structure and store knowledge from BIM standards, enabling rule-based systems and machine learning to be used for compliance checking \citep{zhu2025research}. These methods correlate information, allow for information reuse, and fully express the constraints between building entities. The use of deep learning for pre-classifying regulatory texts can improve the accuracy of structured information extraction by LLMs \citep{chen2024automated}.

Current research also includes addressing the complexity of translating natural language rules, dealing with spatial and geometric relationships, and improving the transparency and usability of these systems \citep{zhao2023compliance}. There is also a push to create digital libraries of rule sets that can be shared online, subdivided by geographical location, which would unite the controls that must comply with a specific regulation \citep{andrich2022check}. The development of an open format for these rule sets could guarantee interoperability between model-checking software. Many studies are now focusing on the integration of LLMs, deep learning models, and ontology knowledge models to improve the efficiency and accuracy of compliance checks \citep{chen2024automated}. Also, there is a growing interest in making these systems more user-friendly and capable of enhancing efficiency and compliance in BIM modeling and procurement processes \citep{de2024enriching}.

\subsection{Current automated tools and their limitations}

Current ACC systems face several challenges, primarily stemming from the complexities of translating natural language regulations into machine-readable rules and the difficulties in ensuring the completeness and accuracy of BIM \citep{lee2023high}. One significant challenge is the difficulty in translating building code text into machine-readable rules \citep{nawari2019blockchain}. Building codes are written in natural language, which is not easily interpreted by computers. The language used in building codes can be ambiguous and complex, with cross-references and inconsistent relationship displays, which increases the complexity of interpreting the sentences and creates discrepancies \citep{preidel2015automated}. Moreover, building codes often have national, regional, and cultural variations in wording and application, making it difficult to develop a universally applicable system \citep{lee2023high}. This issue is further compounded by the fact that building codes are updated frequently, requiring constant modifications to the automated systems \citep{anderson2020using}. The lack of a standardized format for building codes and the absence of a unified conclusion in the ACC domain also contribute to interoperability issues between BIM and ACC systems \citep{altintacs2022integrating}.

Another major challenge is the need for extensive preprocessing and preparation of BIM models for checking \citep{bloch2020clustering}. Existing platforms often require users to manually supplement missing information, correct inaccuracies, or address incomplete data in the model before checking can begin. This process, known as normalization, is labor-intensive, time-consuming, and prone to errors. Furthermore, the complexity of building designs increases the difficulty of ensuring that all necessary information is included and accurate in the model \citep{ismail2017review}. The lack of clearly defined information requirements and variations in modeling practices also contribute to inconsistencies and errors in BIM data \citep{de2024enriching}. Moreover, many systems are too focused on specific domains, such as particular building codes or safety regulations, which limits their general applicability and integration with other systems \citep{fan2019rule}.

% To address these challenges, several software tools now integrate with Revit to automate building code compliance checking, particularly for standards like the IBC. For example, UpCodes AI offers real-time in-model checking using AI, flagging violations related to stairs, doors, ramps, and clearances while linking them to the relevant code sections \citep{upcodes2025}. SMARTreview provides deep IBC analysis through its Revit plugin and produces formal compliance reports (CPR) accepted by some city permitting departments. Solibri, though external, offers extensive rule-based checking (e.g., accessibility, egress) after importing Revit models via IFC. Autodesk’s own Model Checker is a free, configurable tool within Revit, which users can adapt for custom code rules. Bureau Veritas’s icheck focuses on California’s accessibility code (CBC Chapter 11B), while EvolveLAB’s Revit Code Tools automate occupancy, egress, and plumbing calculations through Revit schedules and tags.

To address these challenges, several software tools have been developed to automate compliance checking directly within BIM environments, particularly Revit. For instance, UpCodes AI provides real-time in-model checking using AI, flagging violations related to stairs, doors, ramps, and clearances while linking them to the relevant code sections \citep{upcodes2025}. SMARTreview offers in-depth analysis of the International Building Code (IBC) through its Revit plugin and generates formal compliance reports accepted by some city permitting departments \citep{smartreview2025}. Solibri, while external, allows for extensive rule-based checking, including accessibility and egress, by importing Revit models through IFC \citep{solibri2025}. Autodesk's own Model Checker is a free, configurable tool within Revit, enabling users to create custom code rules \citep{autodeskmodelchecker2025}. icheck focuses on California’s accessibility code (CBC Chapter 11B) \citep{icheck2025}, and EvolveLAB’s Revit Code Tools automate occupancy, egress, and plumbing calculations using Revit schedules and tags \citep{evolvelab2025}.

% These tools vary in scope and sophistication—some check for clear pass/fail conditions while others assist with calculations and documentation. Adoption is growing in the AEC industry, with firms using them to catch issues early and reduce back-and-forth during plan reviews. While no tool fully replaces professional judgment or covers every nuance of the codes, they significantly improve design efficiency and baseline code compliance. Among these, UpCodes and SMARTreview are strong choices for direct compliance checking inside Revit, while Solibri remains a robust external option for broader model quality control.

Although these tools offer varying levels of automation and sophistication, they are not without limitations. Many of them rely on hard-coded rules, which restrict flexibility, making it difficult to adapt to new or changing regulations. Others may struggle with complex geometric relationships, require extensive manual configuration, or are constrained by their focus on specific code standards. These limitations reduce their ability to dynamically adapt to diverse regulatory requirements and complex BIM models.

Furthermore, many existing ACC systems also suffer from a lack of transparency and flexibility \citep{preidel2015automated}. Hard-coded rules, while enabling the checking of specific provisions, are difficult to maintain, modify, and scale \citep{anderson2020using}. Users are often limited to predefined checking capabilities and cannot customize or adjust the rules to meet their specific project needs \citep{lee2015validations}. This lack of user involvement and transparency reduces the acceptance of these systems among domain experts \citep{borrmann2018building}. On the top of that, current systems may lack the ability to check complex, geometric relationships between components \citep{wang2023automated}.

\subsection{Applications of Large Language Models in AEC}
LLMs represent a major advancement in artificial intelligence (AI), showcasing transformative potential across various sectors, including the AEC industry \citep{anderson2020using}. These models, built on transformer architectures and trained on extensive datasets, excel at complex language-related tasks such as translation, summarization, and content generation \citep{minaee2024large}. In construction management, LLMs have been increasingly utilized to automate tasks like translating building regulations into computable formats and integrating regulatory requirements into compliance systems. For example, \citet{fuchs2024using} demonstrated how GPT-3.5 and GPT-4 could structure regulatory texts into machine-readable formats, while \citet{zhang2023can} showed that ChatGPT could generate Python code from regulations, showcasing the scalability of LLMs in compliance automation.

A key strength of LLMs lies in their robust natural language understanding and generation, enabled by pre-training on diverse and large-scale datasets. This equips them with the ability to identify complex patterns in language and adapt to downstream tasks with minimal fine-tuning through emergent capabilities like in-context and few-shot learning \citep{chen2024automated}. These features make LLMs particularly suitable for automating resource-intensive tasks in the AEC industry, such as ACC. Traditionally, ACC has been labor-intensive, requiring manual analysis of complex and ambiguous regulatory texts. LLMs streamline this process by extracting structured information, improving accuracy, and reducing errors \citep{fuchs2024using}.

Generative AI models like the GPT series have also demonstrated potential in architectural design through integration with BIM tools. \citet{jang2022interactive} showcased their use as design assistants, interpreting user inputs, suggesting materials, and updating BIM models with AI-driven recommendations. Similarly, \citet{he2025generative} developed a framework using LLMs and a Physics-Based Conditional Diffusion Model (PCDM) to optimize structural designs, such as shear wall layouts, based on real-world conditions like seismic intensities and building heights.

By reducing dependence on large, annotated datasets and extensive manual feature engineering, LLMs overcome limitations of traditional methods, offering a more efficient and scalable approach to automation \citep{fuchs2024using}. They also enhance the compliance process by employing pre-classification techniques and leveraging deep learning models to handle nested conditional statements and ambiguous regulatory language. These advancements highlight their potential to improve productivity, accuracy, and the overall quality of construction processes \citep{chen2024automated}.

However, the application of LLMs in ACC is not without challenges. LLMs can still have difficulty handling highly complex regulatory texts with intricate structures, nested clauses, and conditional statements \citep{chen2024automated}. Furthermore, some studies have primarily focused on simpler regulatory texts, suggesting that more advanced strategies may be necessary to effectively process more complex information. Despite these limitations, LLMs offer promising solutions for automating the interpretation of regulatory texts, enhancing the efficiency and accuracy of compliance checking in the construction sector \citep{zhang2023can}. The ability to translate natural language into computable formats is a key advantage that LLMs bring to the AEC-FM field.

\subsection{Research Gap}

Despite previous studies' efforts to apply LLMs to ACC, several challenges persist in the field. There is still a lack of integration of compliance-checking models directly within the BIM environment, making it difficult to visually recognize compliance and its implications. Besides, there are limitations in utilizing certain types of regulatory documents, as many commercial rule-checking platforms are restricted to specific country- or state-based regulations and are not customizable to meet diverse needs. Existing systems often struggle to dynamically update compliance-checking processes when regulations change. To address these challenges, we investigated an LLM-based approach that can automatically convert regulations into executable Python scripts within a BIM environment.

\section{Methodology}

\begin{figure}[H]
    \centering
    \includegraphics[width=0.75\linewidth]{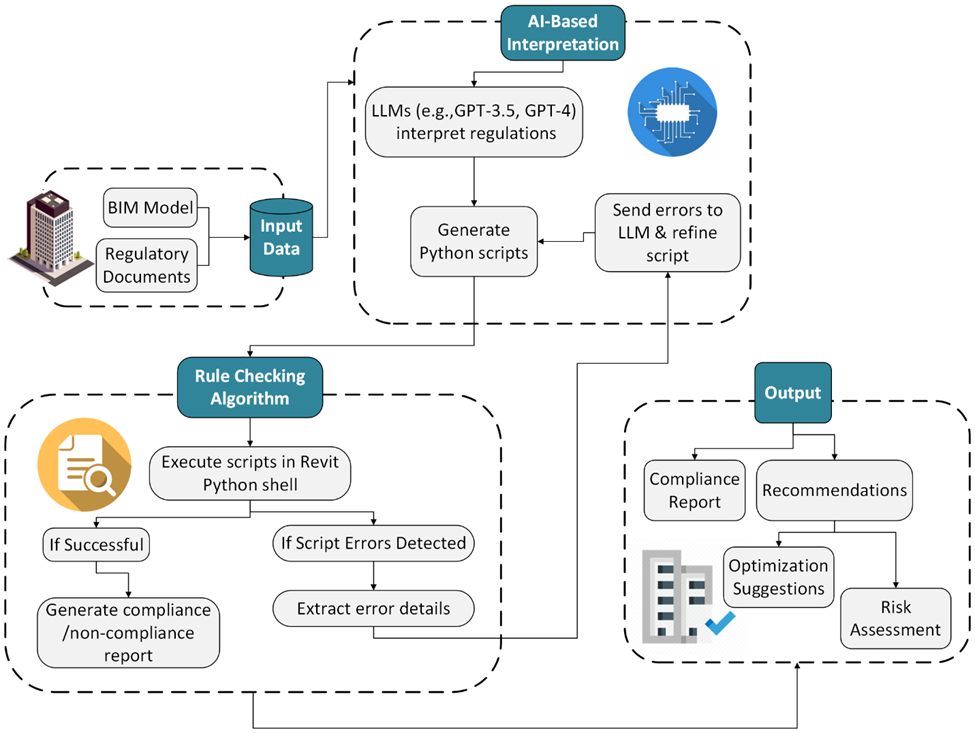}
    \caption{LLM-Based Framework for Rule-Checking in BIM}
    \label{fig:overview}
\end{figure}

Figure \ref{fig:overview} illustrates an LLM-based framework designed to streamline and enhance compliance verification within BIM platforms by leveraging the capabilities of Large Language Models (LLMs), such as GPT-3.5 and GPT-4. The semi-automated framework is divided into four interconnected components: Input Data, AI-Based Interpretation, Rule-Checking Algorithm, and Output. 

The process begins with the input data phase, which integrates two primary sources of information: the BIM model and regulatory documents. The BIM model provides a detailed digital representation of the building/structure’s design, including geometry, spatial relationships, construction elements, and associated metadata. Regulatory documents, on the other hand, include the legal and technical compliance standards that buildings must adhere to. These documents define rules, such as fire safety, accessibility, and structural integrity requirements, which the framework must validate against the BIM model. 

The AI-based interpretation phase is the core of the framework, utilizing the NLP capabilities of LLMs such as GPT, Claude, Gemini, and Llama. During this phase, the LLMs analyze and interpret the regulatory documents. These models are capable of processing complex and unstructured textual information, converting human-readable regulatory standards into machine-readable logic. Based on the interpreted regulations, the LLMs generate Python scripts tailored for automated rule-checking. These scripts are executed within the Revit environment. If errors are produced in executing the generated scripts, the framework sends the error details back to the LLM for refinement. This feedback mechanism ensures the final scripts are functional and reduce the need for manual intervention. 

The rule-checking algorithm phase executes the Python scripts generated in the previous phase to evaluate compliance. The scripts are executed within the Python shell of Revit. If the script execution encounters errors, such as syntax issues, missing data, or logical inconsistencies, these errors are extracted and sent back to the AI-based interpretation phase for refinement. When the scripts execute successfully, the framework generates detailed compliance or non-compliance reports. These reports highlight areas where the BIM model meets the regulations and identify deviations or violations. 

The output phase consolidates and presents the results of the rule-checking process. For compliant models, the framework generates a comprehensive report confirming adherence to all relevant regulations. For non-compliant models, it provides targeted recommendations to address the identified issues. These recommendations are actionable and guide users toward necessary modifications. Beyond basic compliance, the framework also offers additional insights, such as optimization suggestions for improving the BIM model’s design and risk assessments to warn about potential issues related to non-compliance. 

The framework incorporates a feedback that connects the rule-checking algorithm and AI-based interpretation phases. This loop ensures continuous improvement of the Python scripts. By combining the interpretive power of LLMs with the computational capabilities of BIM tools like Revit, the framework will reduce the time and effort required for compliance checks while enhancing accuracy and reliability. 

\section{Case Study}

In this research, we conducted two case studies: one focused on a single-family residential project and the other on an office building.

\subsection{Sample Rules}

In the case studies, 12 sample rules were selected from sources such as the International Residential Code (IRC) \citep{ICC_IRC2021}, International Mechanical Code (IMC) \citep{ICC_IMC2021} and other sources. These rules were converted into Python scripts, as shown in Table \ref{table:rules}. These rules addresses dimensions, object relationships, materials, specific features and fixtures, structural elements, and mechanical components. 
% IRC and IMC were selected for their comprehensive coverage of building, plumbing, mechanical, and electrical requirements, making it ideal for residential projects. Its consistency across locations ensures clear, uniform rules, and its widespread adoption in the United States, along with regular updates for new technologies, makes it a practical choice for this study.

\begin{table}[H]
\small
\centering
\caption{Building Code Rules and References}
\resizebox{\textwidth}{!}{%
\begin{tabular}{@{}p{1cm} p{7cm} p{2.5cm} p{2.5cm}@{}}
\toprule
\textbf{Rule No.} & \textbf{Rule Description} & \textbf{Existing Revit Data} & \textbf{Reference} \\ 
\midrule
1 & The minimum width of the required exit is 36 inches (914 mm), with a net clear width of 32 inches (813 mm). The minimum height of a required exit is 6 feet 8 inches (2032 mm). & Properties of doors & IRC Section R311.2.1, Page 66 \\ 
2 & The minimum clear width of stairways shall be 36 inches. & Stair families & IRC Section R311.7.1, Page 68 \\ 
3 & Porches, balconies, ramps, or raised floor surfaces located more than 30 inches above the floor or grade shall have guards not less than 36 inches in height. & Properties of the guardrail families & IRC Section R312.1.1, Page 72 \\ 
4 & Habitable spaces, hallways, and portions of basements containing these spaces shall have a ceiling height of not less than 7 feet. & Floor levels & IRC Section R305.1, Page 56 \\ 
5 & The window-to-wall ratio in buildings shall not exceed 25\% as stipulated by building code regulations. The ratio is influenced by energy efficiency standards, which might be covered under different codes or local amendments. & IRC Wall area, Window area & IRC  Section R303 \\ 
6 & Every dwelling unit shall have at least one habitable room with not less than 120 square feet of gross floor area. Each additional habitable room, except kitchens, shall have a floor area of not less than 70 square feet. & Room tag & IRC Section R304.1 \\ 
7 & The IRC 2021 Section R307.2 requires a minimum clear space of 21 inches (533 mm) in front of water closets, lavatories, and bidets. & Bathroom fixtures & IRC Section R307.2 \\ 
8 & Toilet Facilities: Every dwelling unit must have a water closet, lavatory, bathtub, or shower. & Plumbing fixture schedules & IRC  Section R306.1 \\ 
9 & Kitchen Requirements: Each dwelling unit must have a kitchen area with a sink. & Room tag \& fixtures & IRC Section R306.2 \\ 
10 & Requirements for wood structural panels used in floor construction. It details material specifications and installation guidelines to ensure floor assemblies meet structural and fire safety requirements. & Material specifications & IRC Section R503.2.4 \\ 
11 & The edge-to-edge distance between any two footings must be at least equal to the width of the larger footing between them. & Footings & \citep{bowles1996foundation} \\ 
12 & Minimum outdoor air ventilation rate required for occupied indoor spaces. For Office Spaces (Business Occupancy), the code states: Each room must receive outdoor air at a rate of: 5 CFM per person (people-based) 0.06 CFM per ft² of floor area (area-based). Total Minimum Ventilation = (5 × occupants) + (0.06 × floor area) & Mechanical Ducts & IMC 2021, Table 403.3.1.1 \\ 
\bottomrule
\end{tabular}%
}
\label{table:rules}
\end{table}

\subsection{Prompt Engineering}

\begin{figure}[H]
    \centering
    \includegraphics[width=0.75\linewidth]{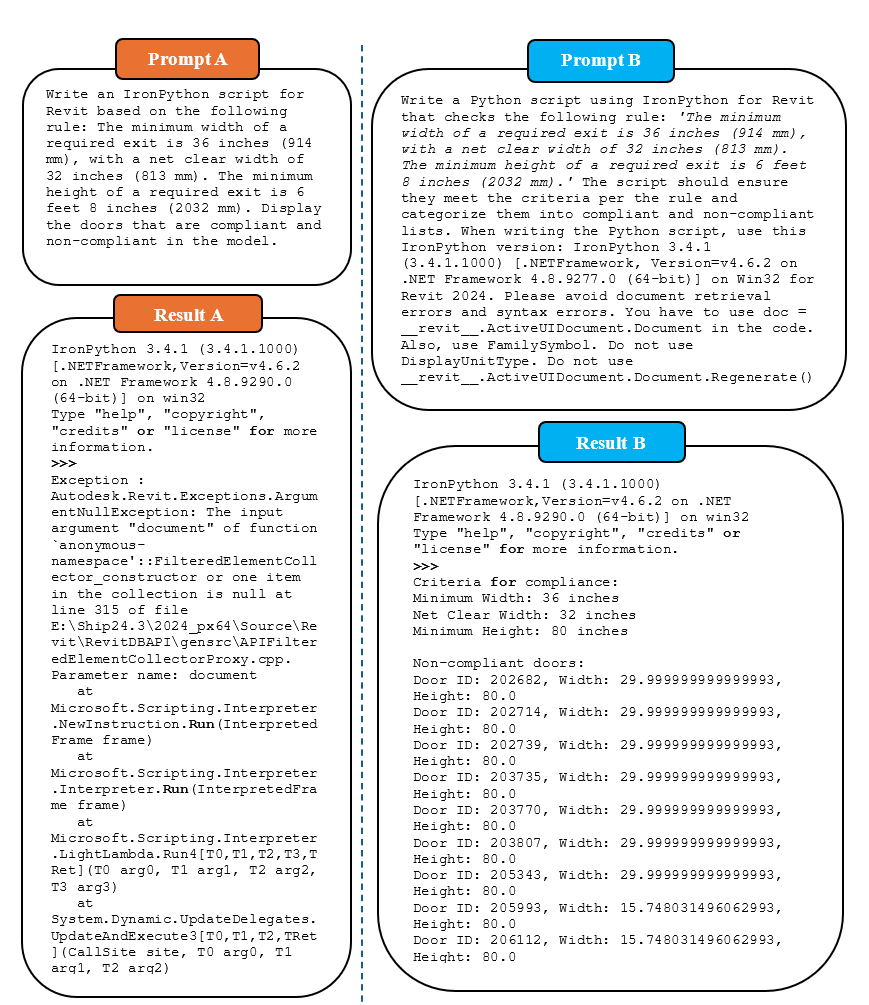}
    \caption{Example Showcasing the Result of Prompt A \& Prompt B on LLM Generated Python Script}
    \label{fig:prompts}
\end{figure}

As part of this research, we developed prompt engineering techniques to enhance the effectiveness of prompts used with LLMs like GPT-4. Figure \ref{fig:prompts} shows an example of the results of two different prompts using Rule 1 in Table \ref{table:rules} as an example. The initial prompt tested, shown as “Prompt A” in Figure \ref{fig:prompts}, when given to a LLM model, generated a Python script. That script, when executed in Revit’s PythonShell, returned an output with an error message, as shown in Result A in Figure \ref{fig:prompts}. To generate a Python script that works in Revit, we tried a number of approaches, including sending the error message back to the LLM, using different phrasing for the prompts, and instructing the LLM to avoid certain types of errors. After several trial-and-error attempts, we were able to identify the optimized prompt structure that is guaranteed to generate an error-free Python script in the Revit PythonShell environment. The optimized prompt, Prompt B in Figure \ref{fig:prompts}, was then applied to Rule 1 again, and the correct result was returned, as shown in Result B in Figure \ref{fig:prompts}.

The optimized prompt's general structure is shown in Table \ref{table:prompt_components}. It consists of four components: the basic prompt, rule description, general instructions, and rule-specific instructions. The first component is the basic prompt, where we define the task that needs to be performed by the LLM. The second component is the rule description, where we provide the actual rule from the regulatory documents that needs to be verified. The third component is the general instructions, which ensure compatibility in the generated script. These instructions specify the versions of PythonShell and Revit being used to provide detailed technical context for the task. Also, since IronPython doesn’t support f-strings, the instructions state that no f-strings should be used in the script to avoid compatibility issues. To further minimize errors, the general instructions emphasize avoiding syntax mistakes and ensuring the script adheres to Revit's requirements. They also address specific elements such as FamilySymbol and DisplayUnitType, and they guide the LLM to generate code that properly interacts with these components within the Revit environment. The last component, rule-specific instructions, focuses on those rules that need to access specific Revit elements, categories, or data attributes, which differ from rule to rule.

\begin{table}[H]
\centering
\caption{Components of the Prompt for Rule-Based Python Script Generation}
\resizebox{\textwidth}{!}{%
\begin{tabular}{@{}p{5cm} p{10cm}@{}}
\toprule
\textbf{Components of the Prompt} & \textbf{Description} \\ 
\midrule
\textbf{Persona} & "You are an expert Python developer with experience in Revit API." \\ 
\hline
\textbf{Basic Prompt} & "Write a Python script in IronPython for Revit to check the following rule." \\ 
\hline
\textbf{Rule Description} & Write the rule as follows. For example, for Rule 1, this part should be: “The minimum width of a required exit is 36 inches (914 mm), with a net clear width of 32 inches (813 mm). The minimum height of a required exit is 6 feet 8 inches (2032 mm).” \\ 
\hline
\textbf{Context} & "This script will be used within Revit 2024, leveraging IronPython 3.4.1 (.NET Framework 4.6.2 on .NET Framework 4.8.9277.0, 64-bit)." \\ 
\hline
\textbf{General Instructions} & The prompt must also include the following instructions: 
\begin{itemize}
    \item Use the following IronPython version: ironpython 3.4.1 (3.4.1.1000), [.netframework, version=v4.6.2 on .net framework 4.8.9277.0 (64-bit)] on win32.
    \item Use the 2024 version of Revit.
    \item Avoid syntax errors and do not use f-strings.
    \item Avoid document retrieval errors. Use \texttt{doc = \_\_revit\_\_.activeuidocument.document} in the code.
    \item Include the \texttt{familysymbol}.
    \item Do not use \texttt{displayunittype}.
    \item Categorize elements into compliant and non-compliant lists.
\end{itemize} \\ 
\hline
\textbf{Rule-Specific Instructions} & See the details in Table \ref{table:rule_specific_instructions}. These instructions relate to the elements from the rule that need to be checked. \\ 
\hline
\textbf{Format} & \begin{itemize}
    \item The Python script should be clean, well-structured, and follow Pythonic conventions.
    \item Add comments to explain each section of the code.
\end{itemize} \\ 
\hline
\textbf{Audience} & "This script is intended for use by construction professionals with basic Python knowledge, working in Revit environments." \\ 
\bottomrule
\end{tabular}%
}
\label{table:prompt_components}
\end{table}

Table \ref{table:rule_specific_instructions} below provides a set of rule-specific instructions that enable LLMs to generate context-appropriate outputs by focusing on relevant Revit elements (e.g., doors, stairs, windows), categories (e.g., ost\_stairs, ost\_plumbingfixtures), and data attributes (e.g., width, height, material properties). These instructions serve as examples for 10 IRC rules derived from Table \ref{table:rules}. These rule-specific instructions are needed primarily because objects in a BIM model are defined differently by various users, and LLMs require instructions to generate scripts that can accurately extract information from the correct object.

\begin{table}[H]
\footnotesize
\caption{Rule-Specific Instructions for LLMs}
\centering
\begin{tabular}{@{}p{1cm} p{12cm}@{}}
\toprule
\textbf{Rule No.} & \textbf{Rule-Specific Instructions for AI} \\ 
\midrule
1 & Not Applicable \\ 
2 & \texttt{stairs\_collector = FilteredElementCollector(doc).OfClass(Stairs)} \\ 
3 & Use \texttt{OST\_StairsRailing} \\ 
4 & Use elevation levels as reference \\ 
5 & Use \texttt{OST\_Windows} and \texttt{OST\_Walls} \\ 
6 & Use \texttt{OST\_Floors} \\ 
7 & Not Applicable \\ 
8 & Use \texttt{OST\_PlumbingFixtures} \\ 
9 & \texttt{plumbing\_fixtures\_collector =} \\ & \texttt{FilteredElementCollector(doc).OfCategory(BuiltInCategory.OST\_PlumbingFixtures)} \\ 
10 & \texttt{[param.Definition.Name for param in element.Parameters]} \\ 
\bottomrule
\end{tabular}

\label{table:rule_specific_instructions}
\end{table}

\subsection{Case Study 1}

\begin{figure}[H]
    \centering
    \includegraphics[width=0.75\linewidth]{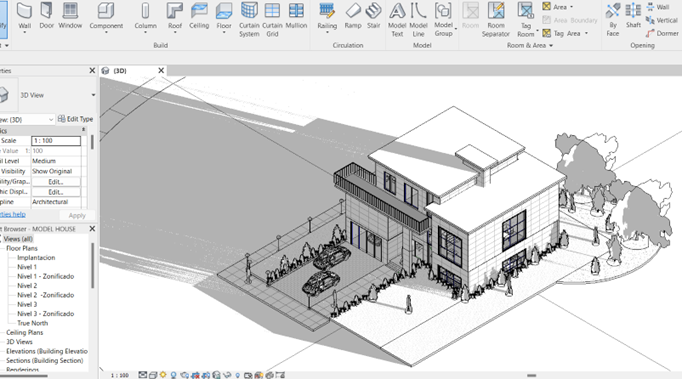}
    \caption{BIM model of a single-family house}
    \label{fig:house}
\end{figure}

A case study of a single-family residential project, shown in Figure \ref{fig:house}, was conducted to demonstrate the effectiveness of the proposed rule-checking framework. The project includes detailed architectural layouts, room dimensions, and refuge areas, providing a suitable dataset for evaluation. Although a small-scale model was used for testing, the compliance-checking process remains consistent across models of any scale.

\begin{figure}[H]
    \centering
    \includegraphics[width=0.75\linewidth]{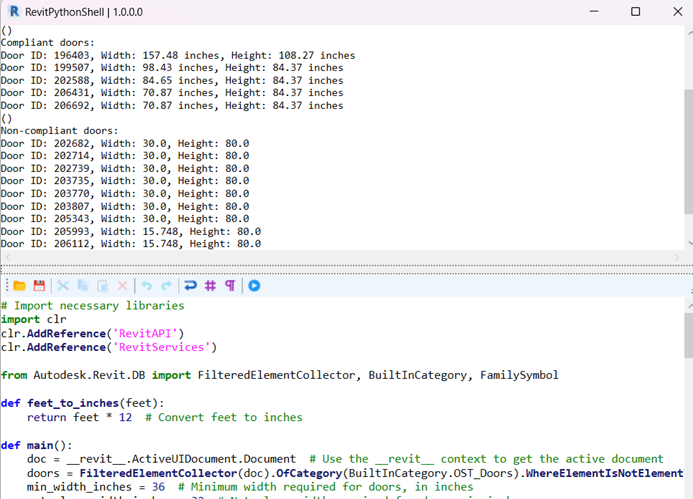}
    \caption{Script displaying both compliant and non-compliant doors}
    \label{fig:door_script}
\end{figure}

% \subsubsection{Results}
Using the optimized prompt structure discussed above, we generated Python scripts for the rules listed in Table \ref{table:rules}. The scripts successfully identified both compliant and non-compliant aspects of the BIM model, and the details of the results are discussed in the section below. Figure \ref{fig:door_script} displays a list of compliant and non-compliant doors for Rule 1 from Table \ref{table:rules}. It includes information about the Door ID along with its width and height dimensions. Doors that meet the minimum width of 36 inches and height of 80 inches are marked as compliant, while those with dimensions below these thresholds, such as a width of 30 inches or less, are classified as non-compliant. Users can utilize the Element ID search in Revit to locate and address the non-compliant elements.

\begin{figure}[H]
    \centering
    \includegraphics[width=0.75\linewidth]{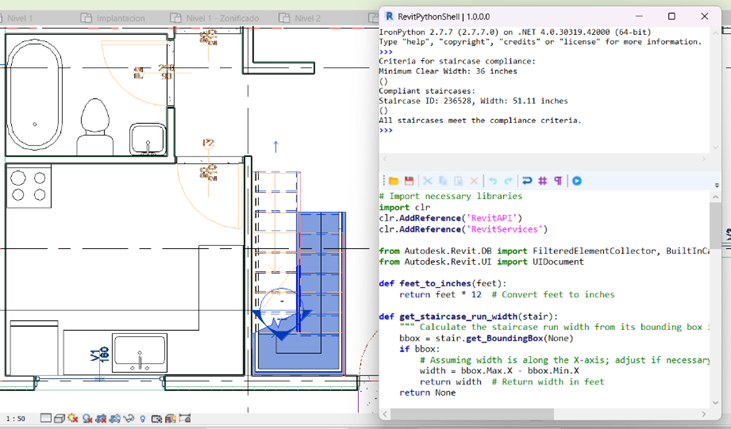}
    \caption{Verifying stairway width compliance with IRC 2021 in Revit using Python script}
    \label{fig:stairway}
\end{figure}

Figure \ref{fig:stairway} measures the clear width of each stairway to verify if it meets the minimum width requirement of 36 inches (914 mm), as specified in Rule 2 from Table \ref{table:rules}. The script extracts the relevant dimensions from the BIM model, converts the units if necessary, and compares them against the IRC requirements. The result demonstrates that the staircase is compliant.

\begin{figure}[H]
    \centering
    \includegraphics[width=0.75\linewidth]{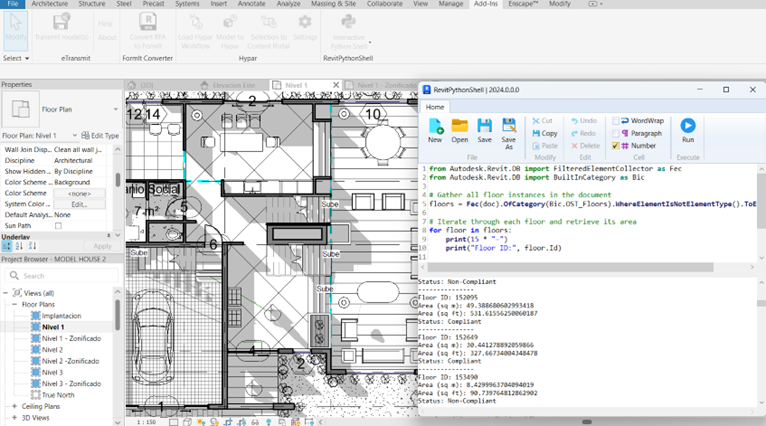}
    \caption{Compliance verification of room size standards using Python integration in Revit.}
    \label{fig:roomsize}
\end{figure}

Figure \ref{fig:roomsize} is based on Rule 6 from Table \ref{table:rules}.The Python script iterates through the habitable rooms, referencing the room tags, and checks whether the rooms have less than 120 square feet of gross floor area. It flags the floor IDs that fall below the minimum standard as non-compliant. For rules like these, the parameters within Revit play a major role. The BIM model must be well-developed with accurate room tags, names, and boundaries to ensure that the correct elements or areas are verified, rather than all rooms.

\begin{figure}[H]
    \centering
    \includegraphics[width=0.75\linewidth]{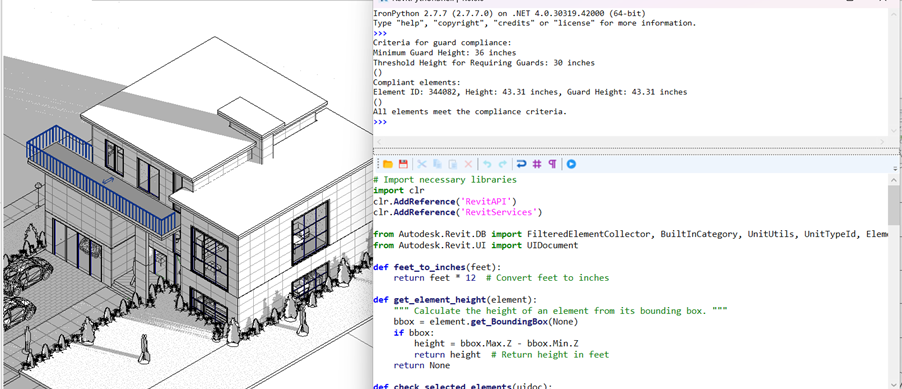}
    \caption{Python script in Revit verifying guardrail height compliance with IRC 2021}
    \label{fig:guardrails}
\end{figure}

Figure \ref{fig:guardrails} demonstrates the results for Rule 3, which requires guardrails on elevated surfaces to be at least 36 inches (914 mm) in height. The script in Figure \ref{fig:guardrails} evaluates each guardrail in the model against this height requirement and flags any non-compliant elements. For instance, a non-compliant guardrail is flagged as Element ID: 654321, with a height of 35 inches and a guard height of 32 inches. Other examples of evaluating relationships between different objects include checking window-to-wall ratio compliance and clear space in front of bathroom fixtures.

\begin{figure}[H]
    \centering
    \includegraphics[width=0.75\linewidth]{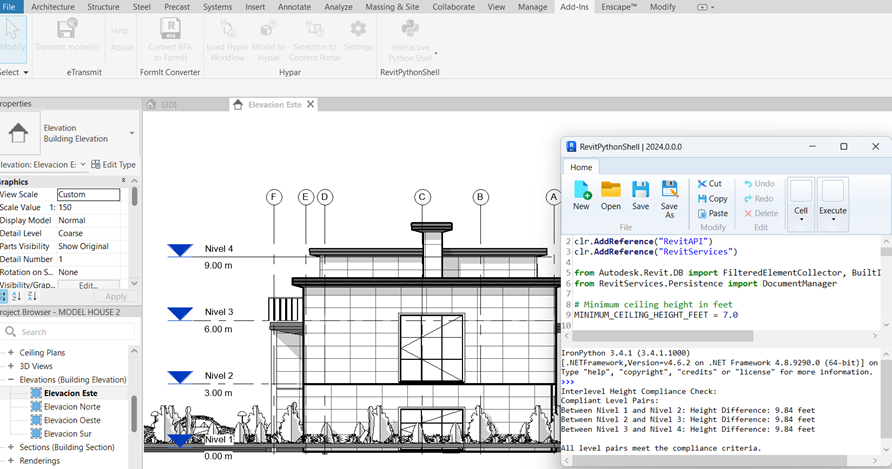}
    \caption{Python script validating ceiling height compliance in Revit}
    \label{fig:ceiling}
\end{figure}

Figure \ref{fig:ceiling} verifies Rule 4 from Table \ref{table:rules}. This rule checks whether all habitable spaces have a ceiling height of at least 7 feet. The output relies on the elevation level parameters to verify the level difference between the floor and ceiling for these rooms. The image below shows that the ceiling height is 9.84 feet and also provides the floor level. The final output confirms that the building is compliant with this rule.

\begin{figure}[H]
    \centering
    \includegraphics[width=0.75\linewidth]{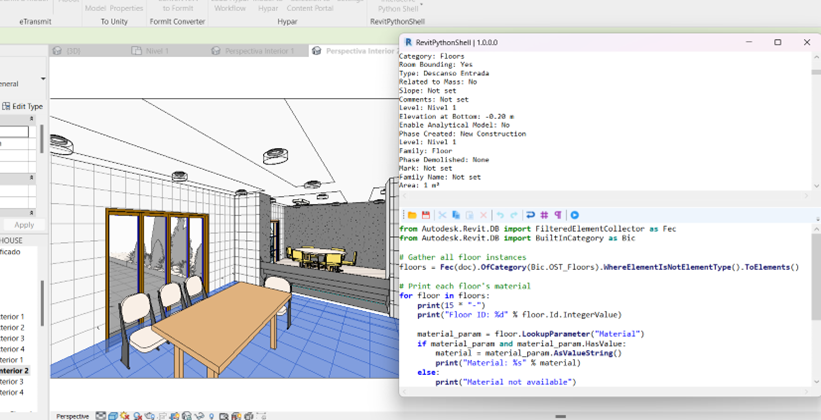}
    \caption{Process of performing compliance check for the Flooring material in Revit}
    \label{fig:flooring}
\end{figure}

Figure \ref{fig:flooring} shows the results for Rule 10, which outlines material specifications and specifies the required thickness and span rating for these panels.  It identifies the room boundary, flooring material, floor level, Element ID, and other details essential for ensuring compliance with floor specifications.

\begin{figure}[H]
    \centering
    \includegraphics[width=0.75\linewidth]{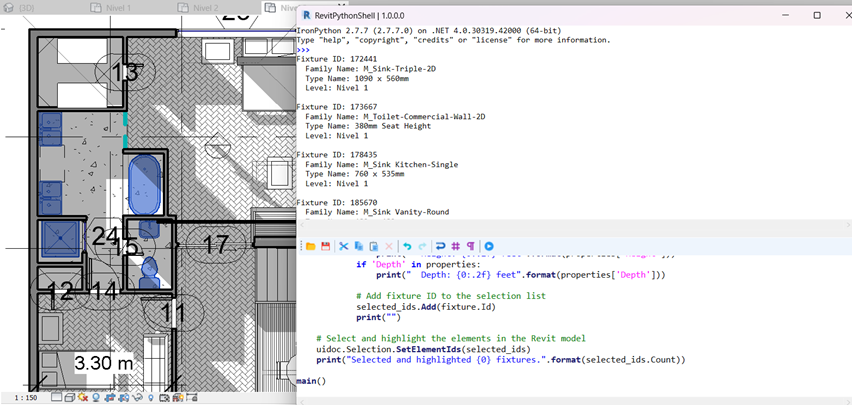}
    \caption{Python script in Revit identifying specific toilet fixtures, detailing their attributes}
    \label{fig:toilet}
\end{figure}

Figure \ref{fig:toilet} shows the compliance output for Rule 8. The Python script interacts with Revit to detect all toilet fixtures and their properties. It verifies the fixture type, placement coordinates, and spatial relationships with nearby items. The output displays toilet fixtures and their components, including water closets (W.C.) and washbasins, along with their design specifications.

\begin{figure}[H]
    \centering
    \includegraphics[width=0.75\linewidth]{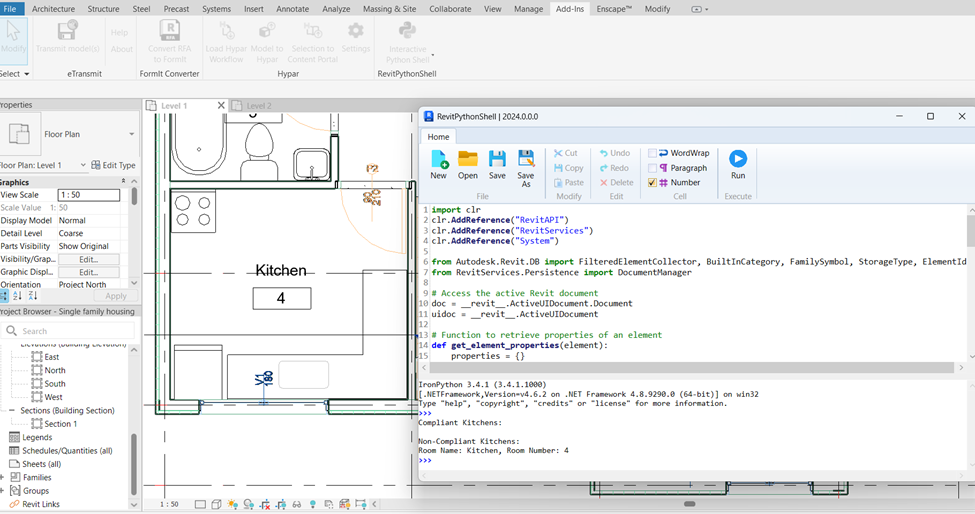}
    \caption{Kitchen compliance check in Revit for required components.}
    \label{fig:kitchen}
\end{figure}

Figure \ref{fig:kitchen} shows the compliance verification output based on Rule 9. In compliance verification, it is critical to check for the presence of required elements in designated spaces. In compliance verification, it is critical to check for the presence of required elements in designated spaces. This process facilitates such checks. The results indicate that the sink is missing from the kitchen, resulting in a non-compliant status. Moreover, the output provides details such as the room number or name, depending on the parameters available in Revit.

\subsection{Case Study 2}

% In the second case study, we apply the proposaed methedology to the office building model. Figure \ref{fig:study2} shows the demonstration of the office building model with the selected 8 rules from Table \ref{table:rules}. While the other rules are the same as in case study 1, Rule 11 demonstrates checking compliance with an rule stating that the edge-to-edge distance between any two footings must be at least equal to the width of the larger footing. Rule 12 defines the minimum outdoor air ventilation rate required for occupied indoor spaces. It shows the compliance check for an IMC code. This rule defines the minimum outdoor air ventilation rate required for occupied indoor spaces, specifically for office spaces (business occupancy). The script first collects all the rooms in the model, then collects all air terminals (such as diffusers). It goes through each diffuser and checks if it has a flow parameter (CFM). If the flow is available, it adds that diffuser's flow to the correct room. After calculating the airflow for each room, the script checks if it meets the area-based requirement (0.06 CFM per square foot of floor area). If the room's airflow is less than the required amount, it is flagged as non-compliant.

\begin{figure}[H]
    \centering
    \subfloat[Rule 2: stairways]{%
        \includegraphics[height=2cm]{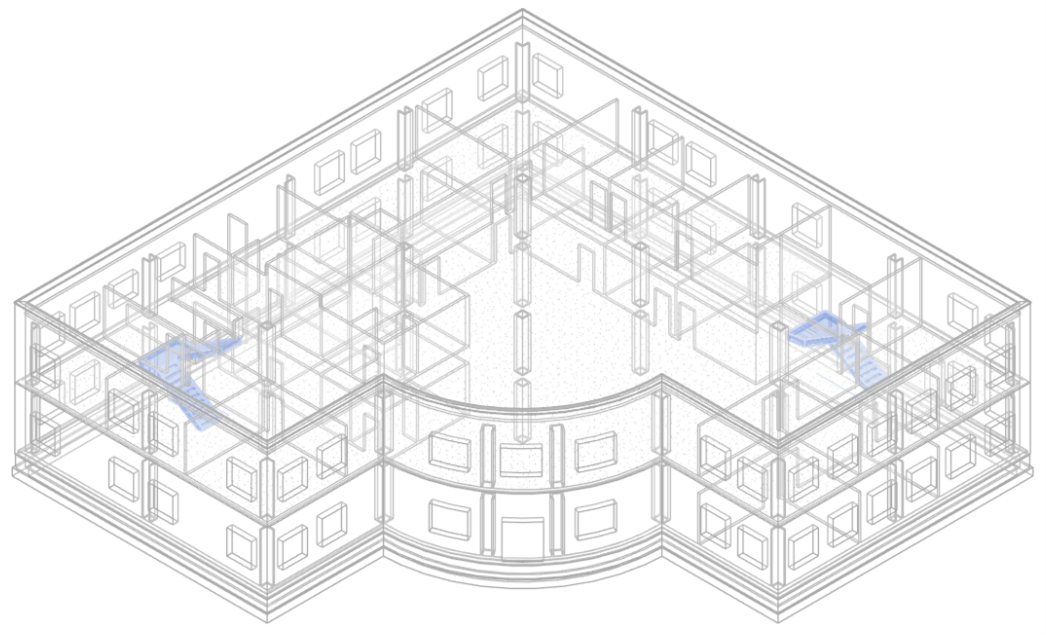}
    }
    % \hfill
    \subfloat[Rule 4: habitable spaces]{%
        \includegraphics[height=2cm]{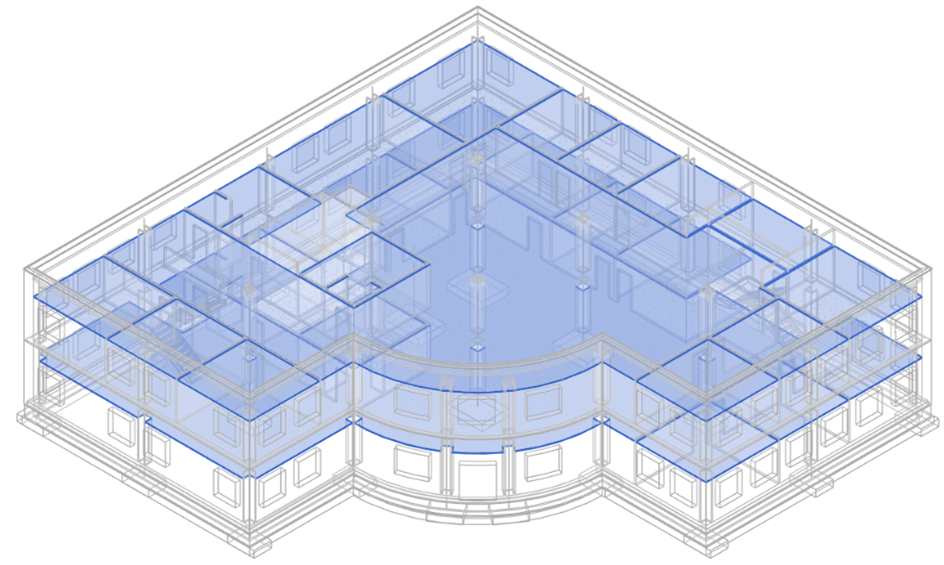}
    }
    % \hfill
    \subfloat[Rule 5: windows-to-wall ratio]{%
        \includegraphics[height=2cm]{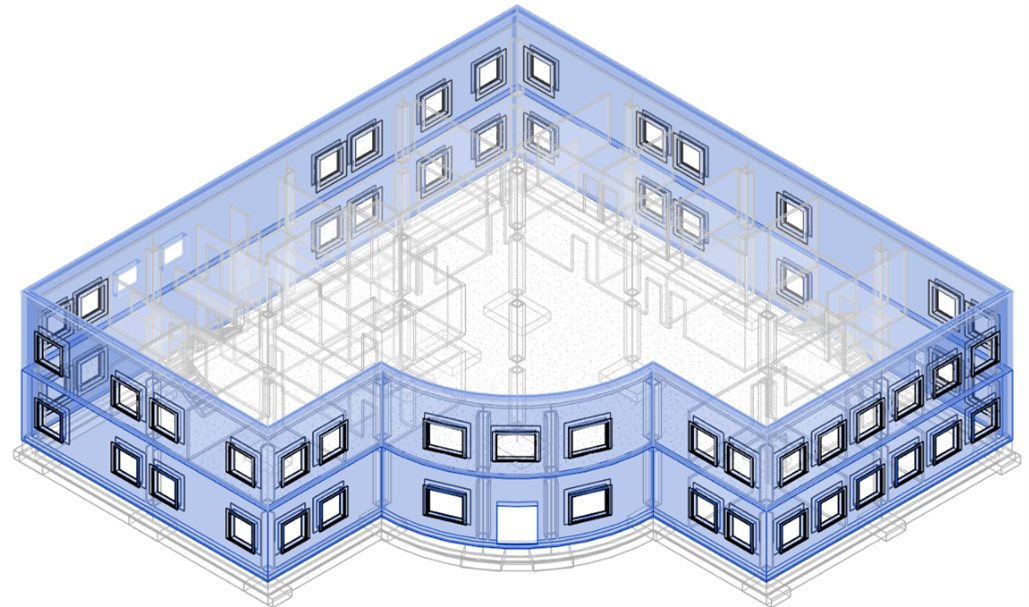}
    }
    \subfloat[Rule 6: floor area]{%
        \includegraphics[height=2cm]{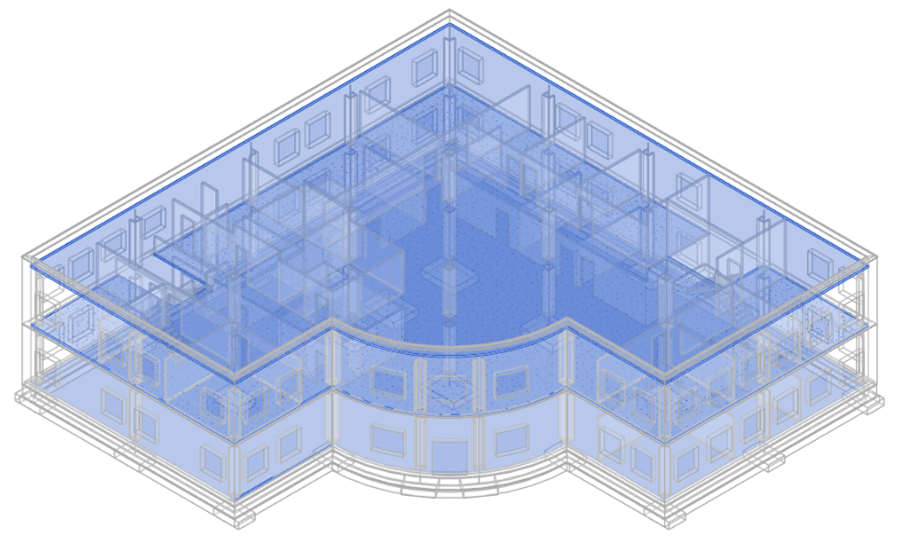}
    }
    % \hfill

% \vspace{0.3cm}  % Adds some vertical spacing
    
    \subfloat[Rule 8: toilet]{%
        \includegraphics[height=2cm]{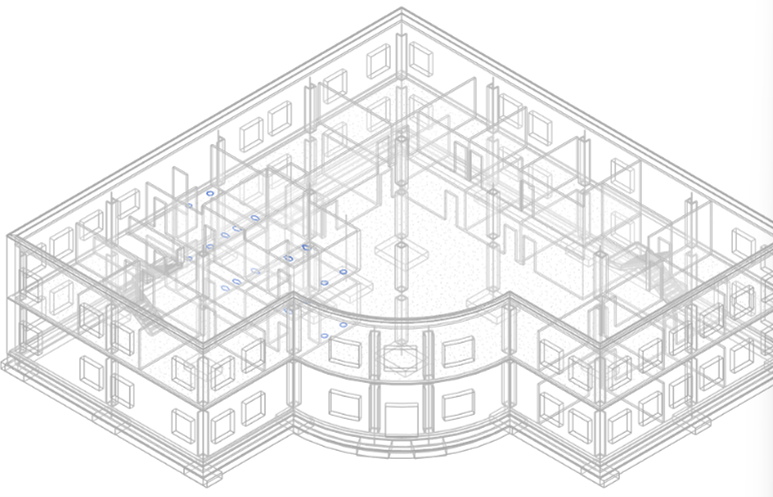}
    }
    % \hfill
    \subfloat[Rule 10: floor]{%
        \includegraphics[height=2cm]{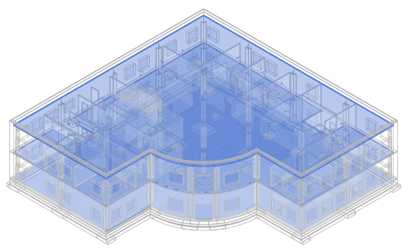}
    }
    \subfloat[Rule 11: edge-to-edge distance]{%
        \includegraphics[height=2cm]{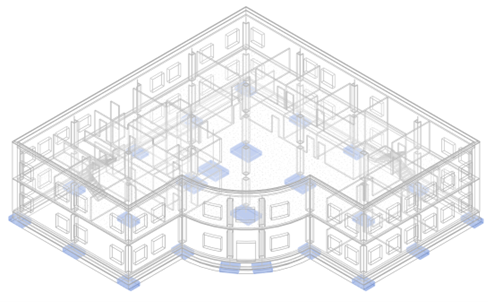}
    }
    \subfloat[Rule 12: ventilation]{%
        \includegraphics[height=2cm]{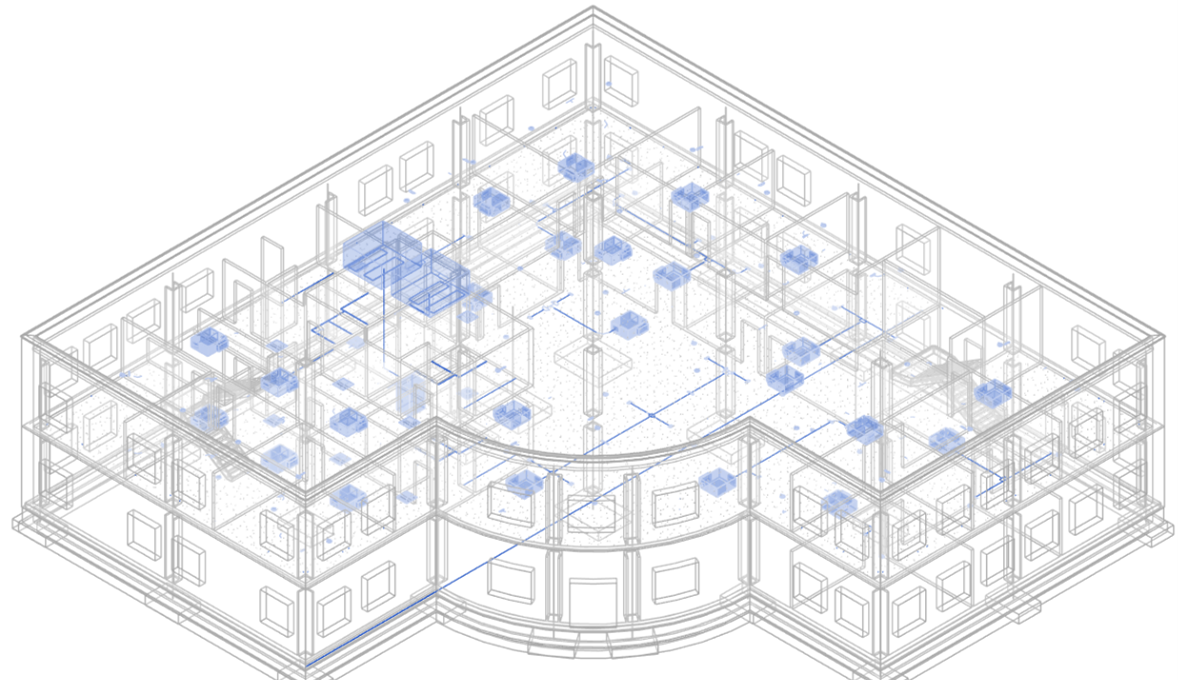}
    }    

    \caption{Case Study 2 Model}
    \label{fig:study2}
\end{figure}

In the second case study, we apply the proposed methodology to the office building model. Figure \ref{fig:study2} illustrates the office building model with the selected 8 rules from Table \ref{table:rules}. While the other rules remain consistent with those in Case Study 1, Rule 11 assesses compliance with the requirement that the edge-to-edge distance between any two footings must be at least equal to the width of the larger footing. Rule 12 defines the minimum outdoor air ventilation rate required for occupied indoor spaces, specifically for office spaces (business occupancy) in accordance with the International Mechanical Code (IMC). The script first identifies all rooms in the model and then collects all air terminals (such as diffusers). It iterates through each diffuser, checking if it has a flow parameter (CFM). If the flow data is available, the script assigns that diffuser's flow to the corresponding room. After calculating the airflow for each room, the script verifies if it meets the area-based requirement (0.06 CFM per square foot of floor area). If a room's airflow is insufficient, it is marked as non-compliant.

\subsection{Generating Compliance Reports Using LLMs' API}
\begin{figure}[H]
    \centering
    \includegraphics[width=0.90\linewidth]{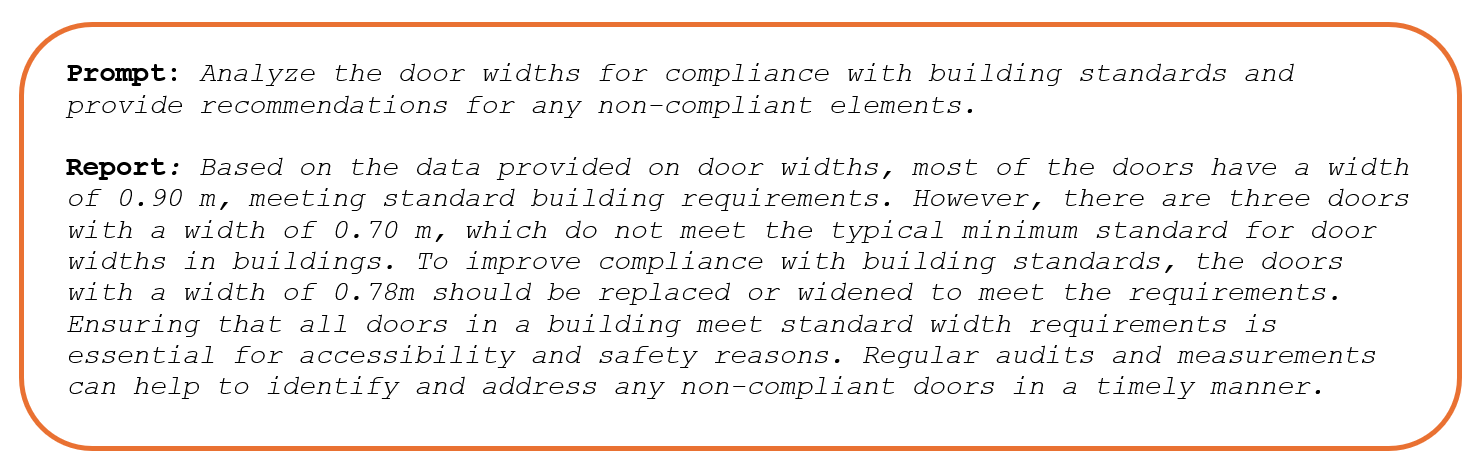}
    \caption{Sample Prompt and Compliance Report for Rule 1}
    \label{fig:sample_prompt}
\end{figure}

This study also utilized LLMs' API to generate compliance reports, as shown in Figure \ref{fig:sample_prompt}, for BIM models by analyzing the output from Revit and providing recommendations for elements that did not meet the standards. The API key was obtained, and the necessary libraries were installed to facilitate the setup. A Python script was developed to extract BIM data. These reports include summaries of key issues, descriptions of non-compliant elements, corrective recommendations, and their impacts on safety and functionality. Figure \ref{fig:sample_prompt} shows a sample prompt and the report generated for Rule 1.

\subsection{Comparison between Different Large Language Models}

Table \ref{table:performance_capabilities} presents the performance of various LLMs using three evaluation metrics: 

\begin{itemize}
    \item Processing Time: The duration (in seconds) required by each LLM to generate Python scripts, as reported by the models themselves.
    \item Correction Attempts: The number of attempts needed for each LLM to produce a functional script without errors. This metric indicates the level of iterative adjustments required.
    \item Status: A binary indicator of whether the LLM successfully generated a functional script on any of its attempts.
\end{itemize}

Moreover, we introduced a quantitative metric, Success Rate, calculated using the following formula:

\begin{equation}
\text{Success Rate (\%)} = \frac{\text{Number of Correctly Executed Scripts}}{\text{Number of Attempts}} \times 100
\label{eq:success_rate}
\end{equation}

This formula measures the LLM's efficiency in producing functional, error-free Python scripts relative to the number of correction attempts made. A higher success rate indicates that the model required fewer attempts to generate a working script.

% Table \ref{table:performance_capabilities} shows the performance of different LLMs in terms of processing time (time taken to generate the scripts, as reported by the LLMs themselves), correction attempts (the number of follow-up prompts required to address errors during script execution), and status (whether the LLM successfully produced a functional script). 

% \begin{equation}
% \text{Accuracy (\%)} = \frac{\text{Number of Correctly Executed Scripts}}{\text{Total Number of Scripts Generated}} \times 100
% \label{eq:accuracy}
% \end{equation}

% Eight LLMs were tested, including ChatGPT 4.0, Claude Sonnet 3.5, Meta Llama 3.1, Microsoft Copilot, Gemini, and Perplexity. Among them, Meta Llama 3.1-405B and Microsoft Copilot failed to generate any successful Python scripts. While Gemini and Perplexity.ai were efficient, they failed to produce correct Python scripts for some rules. In contrast, ChatGPT 4.0 and Claude Sonnet 3.5 emerged as the best models in terms of efficiency. Notably, Claude Sonnet 3.5 proved slightly more effective by reducing the number of correction attempts needed to achieve a successful script.

Eight LLMs were evaluated: ChatGPT 4.0 (\$30/\$60 per 1M tokens), Claude Sonnet 3.5 (\$3/\$15 per 1M tokens), Meta LLaMA 3.1-405B (\$3.50/\$3.50 per 1M tokens), Microsoft Copilot (subscription-based, API price not public), Gemini (\$1.25/\$10 per 1M tokens), Perplexity.AI (\$3/\$15 per 1M tokens), Grok (\$3/\$15 per 1M tokens), and Deepseek (\$0.55/\$2.19 per 1M tokens). Meta LLaMA 3.1-405B and Microsoft Copilot failed to generate any working scripts. Gemini and Perplexity.AI were relatively fast but produced incorrect scripts for several rules, with success rates below 15\%. Deepseek, the lowest-cost option, achieved a moderate 33\% average success rate with few correction attempts. ChatGPT 4.0 and Claude Sonnet 3.5 completed all rule checks; Claude required fewer iterations (3.4 versus 6.8) and delivered a higher average success rate (23.7\%) at roughly one-tenth of ChatGPT's token cost. Grok delivered the best overall performance, reaching a 76.7\% success rate with virtually no retries while maintaining a mid-range price point.

\begin{table}[H]
\centering
\caption{Performance Capabilities of Various LLMs. Success Rate is calculated using Equation~\ref{eq:success_rate}.}
\resizebox{\textwidth}{!}{%
\begin{tabular}{@{}lccccccc@{}}
\toprule
\textbf{Model/LLM (API Price)} & \textbf{Metric} & \textbf{Rule 1} & \textbf{Rule 2} & \textbf{Rule 3} & \textbf{Rule 4} & \textbf{Rule 5} & \textbf{Avg Metrics} \\ 
\midrule
\multirow{4}{*}{ChatGPT 4.0 (\$30/\$60 per 1M tokens)} & Processing Time (s) & 2.0 & 2.0 & 2.0 & 2.0 & 2.0 & 2.0 \\ 
                              & Correction Attempts & 6   & 7   & 9   & 7   & 5   & 6.8 \\ 
                              & Status              & \ding{51} & \ding{51} & \ding{51} & \ding{51} & \ding{51} &    \\ 
                              & Success Rate (\%)   & 14.3 & 12.5 & 10.0 & 12.5 & 16.7 & 13.2 \\ 
\midrule
\multirow{4}{*}{Claude Sonnet 3.5 (\$3/\$15 per 1M tokens)} & Processing Time (s) & 1.5 & 1.5 & 1.5 & 1.5 & 1.5 & 1.5 \\ 
                                    & Correction Attempts & 4   & 2   & 3   & 4   & 4   & 3.4 \\ 
                                    & Status              & \ding{51} & \ding{51} & \ding{51} & \ding{51} & \ding{51} &    \\ 
                                     & Success Rate (\%)   & 20.0 & 33.3 & 25.0 & 20.0 & 20.0 & 23.7 \\
\midrule
\multirow{4}{*}{Meta LLaMA 3.1-405B (\$3.50/\$3.50 per 1M tokens)} & Processing Time (s) & -   & -   & -   & -   & -   & Unknown \\ 
                                      & Correction Attempts & -   & -   & -   & -   & -   & - \\ 
                                      & Status              & \ding{55} & \ding{55} & \ding{55} & \ding{55} & \ding{55} &    \\ 
                                     & Success Rate (\%)   & 0   & 0   & 0   & 0   & 0   & 0 \\ 
\midrule
\multirow{4}{*}{Microsoft Copilot (N/A)} & Processing Time (s) & -   & -   & -   & -   & -   & Unknown \\ 
                                   & Correction Attempts & -   & -   & -   & -   & -   & - \\ 
                                   & Status              & \ding{55} & \ding{55} & \ding{55} & \ding{55} & \ding{55} &    \\ 
                                   & Success Rate (\%)   & 0   & 0   & 0   & 0   & 0   & 0 \\ 
\midrule
\multirow{4}{*}{Gemini (\$1.25/\$10 per 1M tokens)} & Processing Time (s) & 1.5 & 1.5 & -   & 1.5 & -   & 1.5 \\ 
                        & Correction Attempts & 3   & 7   & -   & 8   & -   & 6 \\ 
                        & Status              & \ding{51} & \ding{51} & \ding{55} & \ding{51} & \ding{55} &    \\ 
                        & Success Rate (\%)   & 25.0 & 12.5 & 0   & 11.1 & 0   & 9.7 \\ 
\midrule
\multirow{4}{*}{Perplexity.AI (\$3/\$15 per 1M tokens)} & Processing Time (s) & 3.0 & -   & 3.0 & 3.0 & 3.0 & 3.0 \\ 
                               & Correction Attempts & 3   & -   & 4   & 7   & 6   & 5.0 \\ 
                               & Status              & \ding{51} & \ding{55} & \ding{51} & \ding{51} & \ding{51} &    \\ 
                               & Success Rate (\%)   & 25.0 & 0   & 20.0 & 12.5 & 14.3 & 14.4 \\ 
 \midrule
\multirow{4}{*}{Grok (\$3/\$15 per 1M tokens)} & Processing Time (s) & 1.5 & 1.5   & 1.5 & 1.5 & 1.5 & 1.5 \\ 
                               & Correction Attempts & 0   & 0   & 2   & 0   & 1   & 0.6 \\ 
                               & Status              & \ding{51} & \ding{51} & \ding{51} & \ding{51} & \ding{51} &    \\ 
                               & Success Rate (\%)   & 100 & 100 & 33.3 & 100 & 50 & 76.7 \\     
\midrule
\multirow{4}{*}{Deepseek (\$0.55/\$2.19 per 1M tokens)} & Processing Time (s) & 3.0 & 6.0   & 3.0 & 3.0 & 4.0 & 3.8 \\ 
                               & Correction Attempts & 1   & 3   & 2   & 2   & 3   & 2.2 \\ 
                               & Status              & \ding{51} & \ding{51} & \ding{51} & \ding{51} & \ding{51} &    \\ 
                               & Success Rate (\%)   & 50.0 & 25.0 & 33.3 & 33.3 & 25.0 & 33.3 \\                                    
\bottomrule
\end{tabular}%
}
\label{table:performance_capabilities}
\end{table}

\section{Conclusions}

In this study, a generative AI-based framework was developed to assist construction professionals in evaluating projects for compliance with regulations. The framework leverages large language models (LLMs) to generate Python scripts that can be executed within a Building Information Modeling (BIM) environment. Its effectiveness and reliability were demonstrated through a case study involving a residential building assessed against the IRC building regulations. The framework successfully identified non-compliance issues by flagging problems in the BIM model.

The performance of various LLMs was evaluated in terms of efficiency, speed, and accuracy. Among the models tested, ChatGPT 4.0 and Claude Sonnet 3.5 emerged as the most effective, while Meta Llama 3.1-405B and Microsoft Copilot were unable to generate successful Python scripts, despite multiple attempts.

The study highlights the importance of carefully designing prompts to enable LLMs to generate Python scripts that can execute in Revit without errors. An optimized prompt structure was identified, which includes components such as a basic prompt, rule descriptions, general instructions, and rule-specific instructions. In addition, incorporating a functional script for one specific rule in the prompt was found to aid in generating scripts for other rules.

It was found that the framework was most effective when applied to well-structured BIM models with accurate room tags, boundaries, and naming conventions, as it relies heavily on Revit model properties for verification. Compatibility between software versions also proved crucial; for instance, ensuring that Revit 2024 and PythonShell 2024 are used together is necessary for proper functionality. 

The proposed tool offers significant scientific and technical contributions to the field of construction compliance checking. First, it introduces a novel framework that can be effectively utilized during the pre-construction phase to identify and resolve compliance issues early, reducing risks and costs. Second, the tool's design allows for scalability, making it adaptable to a wide range of construction projects, which demonstrates its versatility and broad applicability. Third, the framework is extendable to incorporate additional structural design regulations, showcasing its flexibility and potential for future enhancements. A key technical innovation is its adaptability to changes in building regulations. Finally, the tool significantly reduces the time required for compliance checking. 

This research has a few limitations that were identified during the case study: (1) it was tested only on two case studies and evaluated only by 12 rules. The optimized prompt structure may require further improvements when tested on a larger set of rules; (2) The testing was conducted on a limited number of Revit versions, which may not fully represent all possible scenarios; (3) The field of LLMs is rapidly evolving, and newer models in the future might perform better than those tested in this study; (4) Although the framework utilizes LLMs to generate Python scripts, the current workflow requires manual intervention for script execution within Revit, as well as for correcting any errors that occur. One potential approach to achieve full automation is the integration of an LLM agent capable of directly interacting with the Revit API. Such an agent could generate, execute, and debug Python scripts in real-time, eliminating the need for manual script transfer. The LLM agent could be designed to receive feedback from Revit, refine the code, and ensure that all rules are correctly applied without user intervention. Future research should explore this approach to further enhance the framework’s automation capabilities; (5) Another limitation of the proposed framework is its reliance on well-structured BIM models with precise metadata. In real-world collaborative environments, BIM models may often contain incomplete, inconsistent, or outdated information, which can negatively impact the framework’s ability to generate accurate compliance checking scripts. Errors or missing metadata can lead to inaccurate interpretations of the model, reducing the effectiveness of automated compliance verification. This dependency on high-quality BIM data may limit the framework's applicability in projects where model quality is inconsistent. Future research should focus on enhancing the framework’s robustness to handle incomplete or inconsistent BIM data. One potential approach is to integrate data validation techniques that automatically identify and correct data gaps within BIM models before initiating compliance checking. Moreover, using LLMs to detect and suggest corrections for missing or erroneous metadata can further improve the framework's resilience; (6) One limitation of the proposed framework is its dependency on the Revit platform, specifically leveraging the Revit API and its unique BIM data structure to automate compliance checking. While this approach is highly effective within Revit, its applicability is currently restricted to this platform. However, the method has the potential to be adapted for other design and modeling tools that support Python scripting and API calls. For example, with appropriate modifications to the script generation component, it could be extended to AutoCAD (for 2D/3D design), ArchiCAD (for BIM), Navisworks (for model coordination), or Rhino/Grasshopper (for parametric design). Future research will explore the applicability of this approach to other platforms to enhance its versatility and practicality.

\bibliographystyle{unsrtnat}
\bibliography{references}

%=====================================
% References, variant B: internal bibliography
%=====================================

% If authors have biography, please use the format below
%\section*{Short Biography of Authors}
%\bio
%{\raisebox{-0.35cm}{\includegraphics[width=3.5cm,height=5.3cm,clip,keepaspectratio]{Definitions/author1.pdf}}}
%{\textbf{Firstname Lastname} Biography of first author}
%
%\bio
%{\raisebox{-0.35cm}{\includegraphics[width=3.5cm,height=5.3cm,clip,keepaspectratio]{Definitions/author2.jpg}}}
%{\textbf{Firstname Lastname} Biography of second author}

% For the MDPI journals use author-date citation, please follow the formatting guidelines on http://www.mdpi.com/authors/references
% To cite two works by the same author: \citeauthor{ref-journal-1a} (\citeyear{ref-journal-1a}, \citeyear{ref-journal-1b}). This produces: Whittaker (1967, 1975)
% To cite two works by the same author with specific pages: \citeauthor{ref-journal-3a} (\citeyear{ref-journal-3a}, p. 328; \citeyear{ref-journal-3b}, p.475). This produces: Wong (1999, p. 328; 2000, p. 475)

%%%%%%%%%%%%%%%%%%%%%%%%%%%%%%%%%%%%%%%%%%
%% for journal Sci
%\reviewreports{\\
%Reviewer 1 comments and authors’ response\\
%Reviewer 2 comments and authors’ response\\
%Reviewer 3 comments and authors’ response
%}
%%%%%%%%%%%%%%%%%%%%%%%%%%%%%%%%%%%%%%%%%%
% \PublishersNote{}
% %\isPreprints{}{% This command is only used for ``preprints''.
% \end{adjustwidth}
%} % If the paper is ``preprints'', please uncomment this parenthesis.
\end{document}